\documentclass[11pt]{article}
\pdfoutput=1
\usepackage{geometry}
\usepackage{graphicx}
\usepackage{amssymb}
\usepackage{epstopdf}
\begin{document}
\title{Chance and necessity in chromosomal gene distributions}
\author{Rutger Hermsen\footnote{FOM Institute for Atomic and Molecular Physics, Kruislaan 407, 1098 SJ, Amsterdam, The Netherlands}, Pieter Rein ten Wolde$^{*}$ and Sarah Teichmann\footnote{MRC Laboratory of Molecular Biology, Hills Road, Cambridge, UK  CB2 2QH}}

\maketitle

\begin{abstract}
{\bf By analyzing the spacing of genes on chromosomes, we find that transcriptional and RNA-processing regulatory sequences outside coding regions leave footprints on the distribution of intergenic distances. Using analogies between genes on chromosomes and one-dimensional gases, we constructed a statistical null model. We have used this to estimate typical upstream and downstream regulatory sequence sizes in various species. Deviations from this model reveal bi-directional transcriptional regulatory regions in \emph{S. cerevisiae} and bi-directional terminators in \emph{E. coli}.}
\end{abstract}

\section{Probability distributions of intergenic distances}

The probability distributions of intergenic distances are shaped by stochastic processes, such as insertions, deletions, inversions and duplications, and by natural selection. Although the former tends to randomize the distribution, the latter introduces biases if there are functional reasons for genes to be spaced in a particular way \cite{Warren:2004:JMB}. Here we compare data of both \emph{Escherichia coli} and different fungal species to a statistical mechanics model to study which features can be explained by random processes only, and which require an explanation in terms of functionality.
 
\section{The Constant-Force model}

 \begin{figure*}[tbp]
   \centering
\includegraphics[width=6cm]{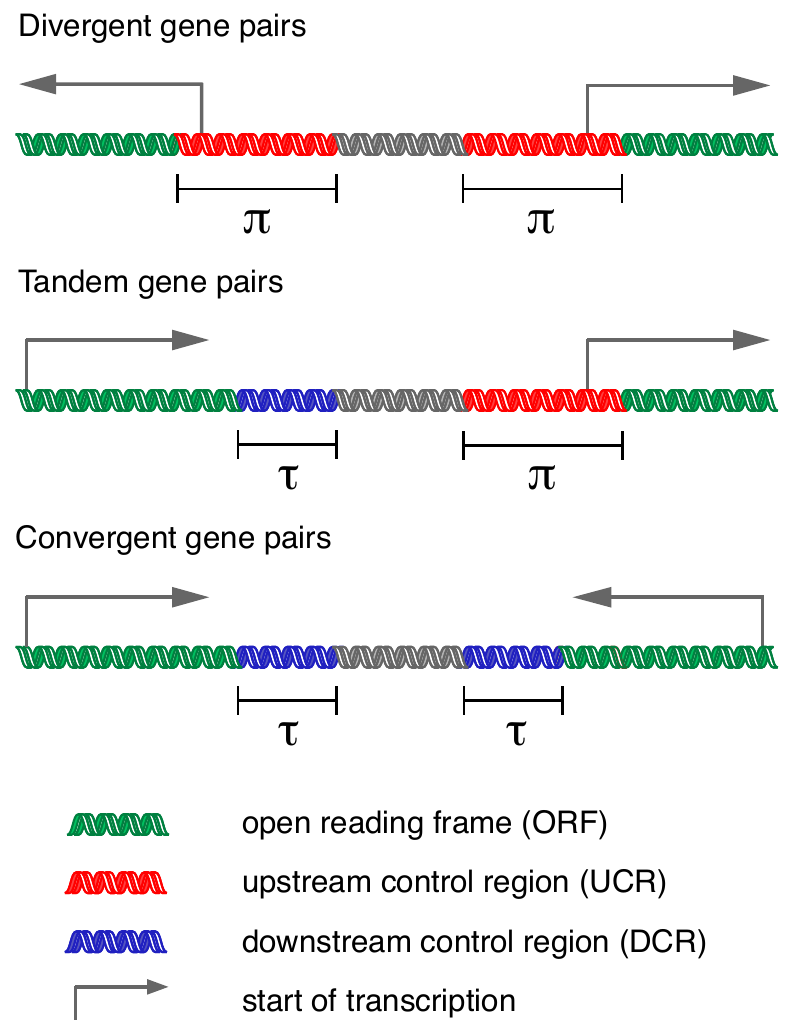}
   \caption{The Constant-Force model. This model assumes that a 5'-UTR, a basal promoter and a \emph{cis}-regulatory region are present upstream of every ORF. We call this the upstream control region (UCR), and assume it has a fixed size $\pi$  . Downstream of each ORF, the 3'-UTR and possibly a transcriptional terminator and RNA processing signals are present, to which we jointly refer as the downstream control region (DCR), assumed to have length $\tau$. The figure shows that ORFs neighboring on the DNA can have three mutual orientations: divergent (D), tandem (T) or convergent (C). This also leads to three kinds of intergenic regions: D regions contain two UCRs, while T regions contain one UCR and one DCR, and C regions have two DCRs. In \emph{S. cerevisiae}, the frequencies of D, T and C regions are 26.3\%, 48.3\% and 25.4\% respectively, which is close to the random proportions 1:2:1. This holds for most fungi. In \emph{E. coli}, T regions are more frequent due to the organization of its genes in operons (17.5\%, 66.7\% and 15,7\%).}
 \label{fig:figure1}
\end{figure*}

The Constant-Force (CF) model is based on two observations (see Fig. \ref{fig:figure1}). First, open reading frames (ORFs) usually do not overlap, even if their density is high. For example, if the genes of \emph{Saccharomyces cervisiae} (budding yeast) were randomly distributed, 78\% of the ORFs would overlap with another ORF, whereas in reality, only 9\% do. 

 \begin{figure*}[tbp]
   \centering
\includegraphics[width=\textwidth]{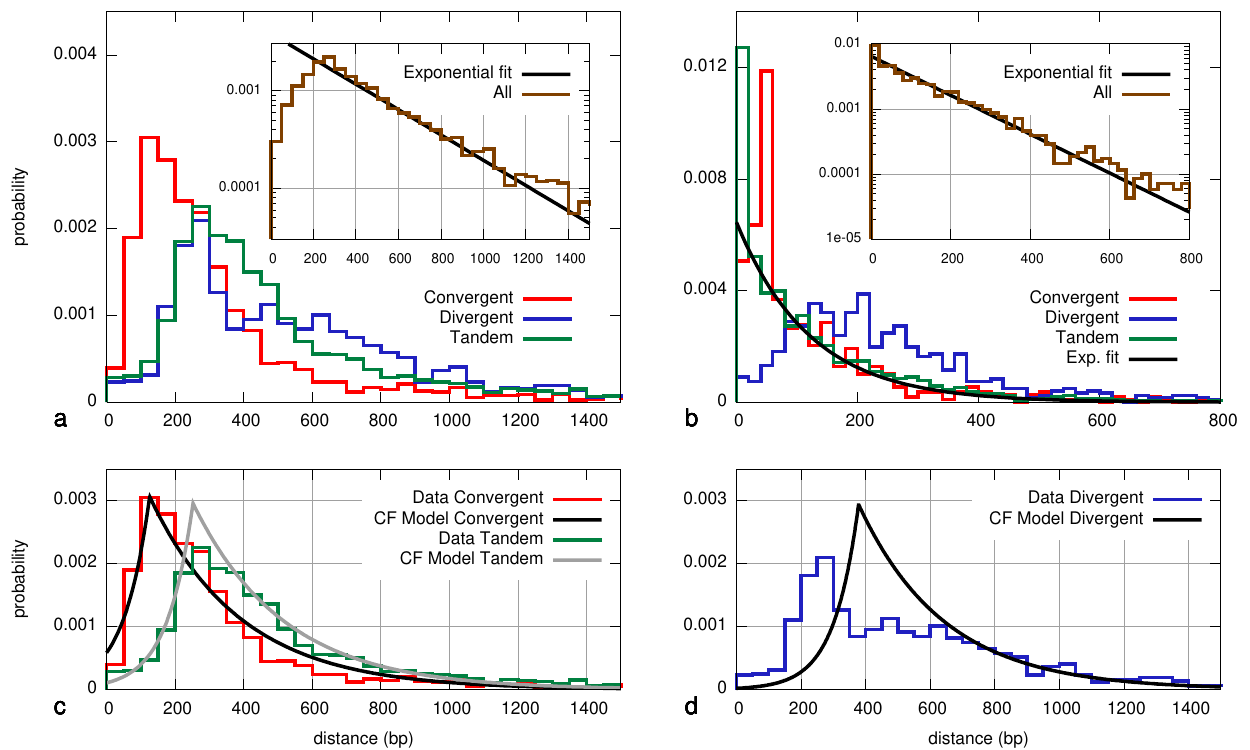}
   \caption{Probability distributions of intergenic distances in \emph{S. cerevisiae} and \emph{E. coli}.
(a) Probability distributions of intergenic regions in \emph{S. cerevisiae}. The distributions of distances between convergent (C), divergent (D) and tandem (T) gene pairs. C intergenic regions are, on average, shorter than T regions; the D regions are longest. Note also that the divergent distribution has a bimodal shape, with a peak at $n\approx275$ bp and one at $n\approx 500$ bp. Inset: the distribution of all intergenic regions is exponential for distances larger than 300 bp (scale parameter: 335 bp), but has a ``dip'' at shorter distances. This dip, we argue, is a footprint of UCRs and DCRs. (b) As panel (a), but for \emph{E. coli}. The T distribution in the main plot is exponential, except for an accumulation in the first bin. This accumulation is the result of intergenic regions inside operons, which are not separated by control regions and therefore can be arbitrarily close together \cite{Lesnik:2001:NAR}. The C distribution is also exponential, except for a peak at 20--60 bp, where \emph{S. cerevisiae} has a ÒdipÓ instead. We predict that this peak is the result of bi-directional terminator sequences. The inset again shows that the distribution of all intergenic regions is largely exponential (scale parameter: 145 bp ). (c) Simultaneous fit of the Constant Force (CF) model to the C and T distributions of \emph{S. cerevisiae}. The model fits the data surprisingly well. (d) The D distribution of \emph{S. cerevisiae} and the expected distribution according to the CF model. Clearly, the bi-modal shape of the data is not consistent with the CF model. We predict that the set of divergent intergenic regions in \emph{S. cerevisiae} consists of two subpopulations: those containing two independent \emph{cis}-regulatory regions, responsible for the second peak, and those containing one bi-directional \emph{cis}-regulatory region.}
 \label{fig:figure2}
\end{figure*}

Second, ORFs are rarely very close together (e.g. see Fig. \ref{fig:figure2}a for \emph{S. cerevisiae}). We hypothesize that this is caused by functional sequences directly upstream and downstream of the ORFs, which we call upstream and downstream control regions (UCRs and DCRs). UCRs include basal promoters, \emph{cis}-regulatory regions and 5' untranslated regions (UTRs); DCRs consist of 3'-UTRs, transcriptional terminators and RNA-processing signals. If ORFs approach each other closely, these regions need either to overlap or to be very short, which makes such configurations less likely. To test this, we divide the intergenic regions into three subsets, called tandem (T), convergent (C) and divergent (D). (See Fig. \ref{fig:figure1}.) Intergenic regions in subset T should contain one DCR and one UCR, whereas C and D intergenic regions contain two DCRs and two UCRs respectively. As UCR sequences are generally longer than DCRs, we expect that D regions are on average longer than T regions and C regions are shortest, which is indeed the case (see Fig. \ref{fig:figure2}a and supplementary Fig. \ref{figureFits}).

These observations inspire the following model. We assume that all ORF configurations are equally probable, except for the following constraints: (i) ORFs do not overlap; (ii) UCRs and DCRs can overlap with each other or with ORFs, but every overlapping base pair (bp) in a particular configuration makes this configuration a factor $q$  less probable. For simplicity, we assume that in a given organism all UCRs and DCRs have a fixed length, $\pi$ and $\tau$ respectively.

This model is equivalent to a one-dimensional system of hard particles with a finite-ranged, repulsive, \emph{constant-force} interaction. Tandem, convergent and divergent ORF pairs interact at a range $\pi + \tau$, $2\tau$  and $2\pi$ , respectively. This mapping enables us to use the formalism of statistical physics to compute the probability distributions corresponding to this model analytically (see the supplementary material).
 
\section{The CF model fits the C and T distribution of \emph{S. cerevisiae}}

The CF model fits the distributions of \emph{S. cerevisiae} convergent and tandem intergenic distances remarkably well (see Fig. \ref{fig:figure2}c). The fit parameters are $\tau=61$ bp for DCRs and $\pi=196$ bp for UCRs. These numbers provide a course estimate of the space required for the transcriptional and translational regulatory signals and RNA processing in \emph{S. cerevisiae}.

Our UCR length prediction of 196 bp is in excellent agreement with the distribution of transcription-factor-binding sites near \emph{S. cerevisiae} start codons, which has its peak at 100-200bp from the start codon \cite{Harbison:2004:N}. Our DCR prediction of 61 bp is supported by bioinformatics analyses of \emph{S. cerevisiae} 3'-RNA processing signals, which show that the majority of these sequences is within 20--90 bp of the stop codon \cite{Helden:2000:NA}. However, Graber \emph{et al.} predict longer 3'-UTRs \cite{Graber:2002:NA} and recent experiments show that the median of 3'-UTRs lengths is $\approx 91$ bp \cite{David:2006:PNAS}, which suggests that our DCR estimate is on the low side. (In the supplementary material, we show that more refined models can provide quantitative agreement).

\section{Typical UCR and DCR sizes for other fungi}

 \begin{table}[htbp]
    \centering
    \begin{tabular}{lccc} 
       Name of organism    & UCR length/bp & DCR length/bp & $q$\\
       \hline
       S. cerevisiae      & $196\pm4$ & $61\pm1$ & $0.985\pm0.001$ \\
      C. glabrata           & $296\pm4$     &  $66\pm2$ & $0.983\pm0.001$\\
       K. lactis       & $295\pm5$  & $38\pm2$ & $0.987\pm0.001$\\
       D. hansenii       & $141\pm4$  & $28\pm2$ & $0.976\pm0.001$\\
       \hline
    \end{tabular}
    \caption{Estimates for the UCR and DCR sizes for various fungal species.
    The errors given in the table are the uncertainties of the fit parameters and as such should not be interpreted as variances of these quantities in the genome.}
    \label{tab:table1}
 \end{table}

We repeated this approach to estimate the UCR and DCR lengths for three additional fungi using only the ORF coordinates as input. (See Table \ref{tab:table1} and Fig. \ref{fig:CDTvarious} in the supplementary material). We found that UCRs are consistently longer than DCRs. The UCR and DCR lengths seem to vary independently of each other, and no dependence on gene density is apparent. Recently, it has been shown that the distribution of Òrigid DNAÓ in \emph{cis}-regulatory regions of fungi correlates with the position of transcription-factor-binding sites \cite{Tirosh:2007:TG}. Our estimates for the UCR lengths correlate well with the position of rigid DNA in these fungi.

\section{\emph{S. cerevisiae} contains many bi-directional UCRs}

We now turn to the spacing of divergent pairs in budding yeast. Interestingly, the corresponding distribution has a bimodal shape (see Fig. \ref{fig:figure2}d) that is even more pronounced in other fungi (Fig. \ref{figureFits} in the supplementary material). The first, narrow peak is centered on  bp; the second peak is broader and is maximal around  bp. This shape is not consistent with the CF model.

Apparently, many divergent intergenic regions are very short: 29\% are $<300$ bp. Because few (10\%, 280 out of 2801) tandem intergenic regions, containing only one UCR, are $<200$ bp, it seems unlikely that two independent UCRs could fit in divergent intergenic regions with a length of the order of 275 bp. Hence we propose that the set of divergent gene pairs is composed of two sub-populations. 

The first population, corresponding to the second peak, consists of pairs of genes that are regulated independently. The other sub-population consists of gene pairs that share a bi-directional \emph{cis}-regulatory region, that is, a regulatory region containing elements such as transcription factor binding sites that regulate the expression of both flanking genes. Such a coupling could force genes to preserve their proximity, thus causing the deviation from the CF model. While bi-directional \emph{cis}-regulatory regions are ubiquitous in \emph{E. coli} \cite{Warren:2004:JMB}, only a few bi-directional UCRs have been reported in \emph{S. cerevisiae} \cite{Bell:1995:CG, Liu:1997:M, Aranda:2006:BBC, Ishida:2006:M}. Based on Fig. \ref{fig:figure2}d, we predict that about 30\% (426 out of 1471) of the divergent pairs are regulated by a shared \emph{cis}-regulatory region. (We list the best candidates in the Supplement.)

If this is true, then one would expect co-expressed divergent pairs to be overrepresented in the first peak rather than the second. This is indeed the case for positively correlated pairs ( $p<0.002$; see Supplement, section \ref{IVA}). Negatively correlated pairs are typically not in the first peak. This contrasts with bi-directional UCRs in bacteria, in which dual regulators often act as a repressor for one of the genes and as an activator for the other, resulting in anti-correlated expression patterns.

We also used Gene Ontology (GO) annotations \cite{Ashburner:2000:G} to test whether the divergent neighbors in the first peak are more often functionally related than those in the second peak. Adopting the method of reference \cite{Resnik:1998:JAIR} to quantify the similarity between GO terms, we indeed found this to be the case ($p < 9\times 10^{-4}$  for biological process,  $p < 5\times 10^{-5}$ for cellular component; see Supplement, section \ref{IVB}).

\section{\emph{E. coli} has many bi-directional terminators}

As mentioned above, bi-directional promoters are well-characterized in \emph{E. coli}. We now show that the distribution of convergent gene pairs in \emph{E. coli} provides evidence for bi-directional transcriptional terminators, which are much less well described.

In accordance with the CF model, the C distribution has an exponential signature (see Fig. \ref{fig:figure2}b). Convergent intergenic regions are expected to contain two DCRs. Given the typical size of Rho-independent terminators ( $\approx 40$ bp) the CF model predicts a dip at short distances ( $< 80$ bp). Instead, there is a significant excess of intergenic regions of size 20 to 60 bp ($p=10^{-13}$ ; see Supplement, section \ref{IVC}). It is unlikely that two terminators would fit into such short intergenic regions.

Rho-independent terminator sequences function by stem-loop formation of the RNA transcript, and hence are largely palindromic. As the complementary strand of a palindromic sequence is a palindrome too, some terminators can function bi-directionally. Indeed, a few bi-directional terminators have been identified experimentally \cite{Postle:1985:C}. Moreover, Lesnik \emph{et al.} used an algorithm called RNAMotif to identify putative terminators and predicted that many of them could function bi-directionally \cite{Lesnik:2001:NAR}. Given that most terminators in \emph{E. coli} start within 60 bp downstream of their ORF \cite{Lesnik:2001:NAR}, genes sharing a bi-directional terminator should usually be close together; this suggests that the peak in the distribution at short distances is caused by bi-directional terminators.

To explain the data, at least 86 bi-directional terminators should be present; this would imply that as many as 23\% of the operons use a bi-directional terminator (see Supplement, section \ref{IVC}).

We tested this, using the data of Lesnik \emph{et al.}\footnote{Although no statistical test is presented, a similar conclusion is reached in ref. \cite{Yachie:2006:FL}}. Indeed, putative terminators that RNAMotif classifies as bi-directional have a tendency to occur in short, convergent intergenic regions, corroborating our hypothesis ($p<0.0003$, see Supplement, sections \ref{IVD} and \ref{IVE}) \cite{Salgado:2000:PNAS}.

\section{Concluding remarks}

The largest limitation of the current model is the assumption that the UCRs and DCRs have fixed sizes. Especially in higher eukaryotes, UCR and DCR lengths often have a high variance; in these cases, it is necessary to include this in the model. In the Supplement we show that this can be done and how more realistic potentials can be chosen.

That being said, the simple CF model describes many universal characteristics of the gene spacing.  It not only quantitatively describes the exponential decay at large distances, but also the ÒrepulsionÓ at short distances due to UCRs and DCRs in prokaryotes and eukaryotes alike. In \emph{E. coli} and fungi, this repulsion provides information about the typical length of UCRs and DCRs using only the ORFs coordinates as input. The model can also serve as a null model for the spacing of genes: deviations from it lead to meaningful predictions about the presence of operons, bi-directional promoters or terminators.

\section{Acknowledgements}

We thank Andre Boorsma and Harmen Bussemaker for providing the yeast expression data. This work is part of the research program of the "Stichting voor Fundamenteel Onderzoek der Materie (FOM)", which is financially supported by the "Nederlandse organisatie voor Wetenschappelijk Onderzoek (NWO)".

\newpage
\appendix

\begin{center}
{\bf \huge Supplementary Text}
\end{center}

\section{The Constant Force Model}

In this section we provide additional information about the Constant-Force (CF) model.

\subsection{Assumptions}\label{IA}

As we explained in the main text, the CF model is based on three assumptions. First, we assume that ORFs cannot overlap. Second, in a given organism, upstream control regions (UCRs)  and downstream control regions (DCRs) have a fixed size ($\pi$ and $\tau$ respectively).  Third, we assume that these control regions can overlap with each other and with nearby ORFs, but that such overlaps are not likely. More precisely, we assume that \emph{whenever a base pair from such a region overlaps with another functional region, be it an ORF, a UCR or a DCR, it makes that particular configuration a factor $q$ less probable}.  For simplicity, we make no distinction between the different kinds of overlap.

A useful analogy can be drawn with a physical system. The proposed model is formally equivalent to a one-dimensional system of hard particles with finite-ranged repulsive interactions. The interaction determined by our assumptions is an interaction with a \emph{constant force}. The range of the interaction depends on the mutual orientation of the neighboring genes. Divergent gene pairs are separated by two UCRs and therefore start interacting at a distance $2\pi$; convergent pairs have two DCRs in their intergenic region and therefore have an interaction range of $2 \tau$, and intergenic regions between tandem pairs contain one DCR and one UCR (interaction range $\pi+\tau$). This analogy allows us to use the formalism of statistical physics to compute the probability distribution of the intergenic distances for this model analytically; the complete derivation follows below.

\subsection{Derivation of the distance distributions:\\ CF interaction with fixed range}\label{IB}

In the CF model described above, the interaction range of the particles depends on their mutual orientation (convergent, divergent or tandem). We first derive the distance distribution for a slightly simpler system, in which the interaction range does not depend on the orientation.

We consider a one-dimensional space (representing the chromosome) of length $L'$ containing $N-1$ particles (representing ORFs). We choose to describe the system in the micro-canonical ensemble, with fixed total energy $E$. The state of the system can be described by a vector $\vec{n}= (n_{1}, n_{2}, \ldots, n_{N})$, where $n_{i}$ is the length of the $i$th inter-particle space. The sum of these numbers, $L\equiv\sum_{i} n_{i}$, is the total free space in the system. The value of $L$ is fixed and $L\gg1$. As the particles occupy part of the total space, $L<L'$.

For now we assume that the particles interact with a finite-ranged CF potential $U(n)$, defined as:
\begin{equation}\label{U}
\frac{U(n)}{kT} = \left \{
\begin{array}{ll}
\epsilon(r-n) & (n < r)\\
0 & (n \geq r),\\
\end{array}
\right.
\end{equation} 
where $r$ is the range of the interaction, and $\epsilon$ is the energy associated with an overlap of one base pair (in units of $kT$); it is related to $q$ as $\epsilon = -\ln(q)$.

In order to compute the probability distribution of intergenic distances, we divide the system into two subsystems. Subsystem 1 (S1) is a particular, but arbitrary, inter-particle space $x$, while subsystem 2 (S2) is the rest of the system. We will compute the probability distribution $P(n_{x})$ of the length $n_{x}$ of space $x$. We define the multiplicity function of subsystem S2, called $\Omega_{2}(L_{2}, E_{2})$, as the number of states accessible for S2 given the available free length for S2, $L_{2}$, and the available energy for S2, $E_{2}$. Note that $L_{2}=L-n_{x}$ and $E_{2}=E-E_{1}=E-U(n_{x})$. Then the probability that inter-particle space $x$ has length $n_{x}$ is proportional to the number of states that are accessible to the rest of the system, S2, given that $x$ has length $n_{x}$:
\begin{equation}
P(n_{x}) \propto \Omega_{2}\left(L-n_{x}, E-U(n_{x})\right).
\end{equation}

By definition, the entropy $\sigma_{2}(L_{2}, E_{2})$ of S2 is the logarithm of $\Omega_{2}(L_{2}, E_{2})$. Therefore,
\begin{equation}\label{plx}
P(n_{x}) \propto e^{\sigma_{2}(L-n_{x}, E-U(n_{x}))}.
\end{equation}
Assuming that $n_{x}$ is small compared to $L$ and that $U(n_{x})$ is small compared to $E$, we can now expand the entropy as follows:
\begin{eqnarray}\label{expans}
\sigma_{2}(L-n_{x}, E-U(n_{x})) =&& \sigma_{2}(L, E) \\
&-& n_{x} \frac{\partial \sigma_{2}(L, E)}{\partial L}\nonumber\\ 
&-& U(n_{x})  \frac{\partial \sigma_{2}(L, E)}{\partial E}\nonumber\\
  &+& \ldots \nonumber
\end{eqnarray}
Note that by the standard Maxwell relations, $\left(\frac{\partial \sigma_{2}}{\partial E}\right)_{L}=1/kT$ and $\left(\frac{\partial \sigma_{2}}{\partial L}\right)_{E}=p/kT$, where $T$ and $p$ are the temperature and the pressure of the system. If $N$ is large, the higher order terms are negligible.

Now we can combine the expansion in equation \ref{expans} with equation \ref{plx} and obtain:
\begin{equation}\label{plxfinal}
P(n_{x}) \propto e^{ -( n_{x}  p + U(n_{x}))/k T}.
\end{equation}
We can calculate this in full using the definition of the potential in equation \ref{U}, arriving at:
\begin{equation}\label{final}
P(n_{x}) = \left \{ \begin{array}{ll}
	c ~e^{-n_{x} \left( \lambda -\epsilon\right)} & (n_{x} < r)\\
	c ~e^{-n_{x} \lambda + \epsilon r}& (n_{x} \geq r).
	\end{array}
	\right .
\end{equation}
Here $\lambda$ is defined as $\lambda\equiv\frac{p}{kT}$.
As we picked inter-particle space $x$ arbitrarily, this probability distribution holds for all inter-particle spaces. The number $c$ is a normalization constant. Given $r$ and $\epsilon$,  the value of $\lambda$ is fixed if we impose the mean inter-particle distance:
\begin{equation}
 \int n_{x} P(n_{x})dn_{x} = L/N.
\end{equation}

Note that, beyond the interaction range, the distribution is exponentially decreasing. Within the interaction range, the distribution is also exponential, but the sign of the exponent depends on the size of $\epsilon$: if the repulsion is strong ($\epsilon>\lambda$), the exponent becomes positive in the interaction range. We also note that if either the range $r$ or the repulsion $\epsilon$ is set to zero, the distance distribution simply becomes a single exponential. The resulting model is known as a Tonks gas~\cite{Tonks:1936:PR}.

\subsection{CF interactions with different ranges}\label{IC}

In the previous subsection we discussed a CF model in which each particle interacts with its neighbors according to one fixed interaction range. In the relevant case, however, the interaction range depends on the mutual orientation of the particles. The interaction potentials for convergent (C), tandem (T) and divergent (D) pairs can be written as follows:
\begin{eqnarray}
\frac{U_{C}(n)}{kT} &=& \left \{
	\begin{array}{ll}
	 \epsilon (2\tau- n) & \textrm{if $n\leq 2 \tau$,}\\
	0 & \textrm{if $n>2 \tau$,}\nonumber
	\end{array}
	\right. \\
\frac{U_{T}(n)}{kT} &=& \left \{
	\begin{array}{ll}
	 \epsilon (\tau + \pi- n) & \textrm{if $n\leq \tau + \pi$,}\\
	0 & \textrm{if $n> \tau + \pi$,}
	\end{array}
	\right. \\
\frac{U_{D}(n)}{kT} &=& \left \{
	\begin{array}{ll}
	 \epsilon (2 \pi- n) & \textrm{if $n\leq 2 \pi$,}\\
	0 & \textrm{if $n> 2 \pi$.}\nonumber
	\end{array}
	\right.
\end{eqnarray}
It is rather straightforward to adjust the calculations in the previous section to this case.

We again divide the system in two parts, S1 and S2, in which S1 consists of one inter-particle region called $x$, and S2 is the rest of the system. The derivation in the previous section applies without alteration up to equation \ref{plxfinal}, irrespective of the orientation corresponding to $x$ (that is: D, T or C). Only in the step from equation \ref{plxfinal} to equation \ref{final}, the difference in the potentials for D, T and C becomes relevant. As a result, the distributions for the D, C and T intergenic regions all have the form of equation \ref{final}, except for a different range $r$, and a different normalization factor $c$:
\begin{eqnarray}
P_{C}(n) &=& \left \{
\begin{array}{ll}
c_{1} e^{-n(\lambda-\epsilon)} & \textrm{if $0\leq n\leq 2 \tau$,} \\
c_{1} e^{-n \lambda+2 \tau \epsilon} & \textrm{if $n > 2 \tau$,}
\end{array}\right.\nonumber\\
P_{T}(n) &=& \left \{
\begin{array}{ll}
c_{2} e^{-n(\lambda-\epsilon)} & \textrm{if $0\leq n\leq \tau+\pi$,} \\
c_{2} e^{-n \lambda + (\tau+\pi)\epsilon}& \textrm{if $n>\tau+\pi$,}
\end{array}\right.\nonumber\\
P_{D}(n) &=& \left \{
\begin{array}{ll}
c_{3} e^{-n(\lambda-\epsilon)} & \textrm{if $0\leq n\leq 2 \pi$,} \\
c_{3} e^{-n \lambda+2\pi \epsilon} & \textrm{if $n>2 \pi$.}
\end{array}\right.
\end{eqnarray}
Here the prefactors $c_{1}$, $c_{2}$ and $c_{3}$ are defined as
\begin{eqnarray}
c_{1} &=& \frac{ \lambda (\lambda-\epsilon)} {\lambda-e^{2 \tau (\epsilon- \lambda)} \epsilon},\nonumber\\
c_{2} &=& \frac{ \lambda (\lambda-\epsilon)} {\lambda-e^{(\tau+\pi) (\epsilon- \lambda)} \epsilon},\\
c_{3} &=& \frac{ \lambda (\lambda-\epsilon)} {\lambda-e^{2 \pi (\epsilon- \lambda)} \epsilon}.\nonumber
\end{eqnarray}

We note that the CF model has four parameters: $\lambda$, $\tau$, $\pi$, and $\epsilon$. However, if we impose the average length of the intergenic regions, this again leads to a constraint that eliminates  one of the parameters. As the total system is a mixture of D, C and T intergenic regions in proportions $f_{D}:f_{C}:f_{T}$ (in most genomes roughly 1:1:2), this constraint becomes:
\begin{equation}\label{average}
 \int n \frac{ f_{D} P_{D}(n) + f_{C} P_{C}(n)+f_{T} P_{T}(n)}{f_{D}+ f_{C}+f_{T}}  dn = L/N.
\end{equation}
We used Monte Carlo simulations to check the validity of these equations and found excellent agreement. 

\section{More detailed models}

The CF model is purposely oversimplified. Such simplified models, with few parameters,  provide insight  into the essential ingredients of the mechanisms studied. At the same time the simplicity of the CF model leads to certain artifacts. Here we show that such artifacts can be alleviated by more detailed models. Below we discuss how one can allow for varying UCR and DCR lengths, and how alternative interaction potentials can be chosen, with distance-dependent forces.

\subsection{Polydisperse UCRs and DCRs}\label{IIA}

The distributions of the CF model have a sharp peak; this is an artifact of our assumption that all UCRs and all DCRs have the same length. We can extend the model to describe systems with varying UCR and DCR lengths.

If UCR and DCR lengths vary, then this results in a varying interaction range $r$. In general, due to differences in the UCR and DCR lengths, the interaction range obeys probability distributions $p_{C}(r)$, $p_{D}(r)$ and $p_{T}(r)$ for the convergent, divergent and tandem intergenic regions respectively. Then at a given pressure $\lambda$ the distributions of intergenic distances are given by
\begin{eqnarray}\label{integrals}
P_{C}(n) &=& \int_{0}^{\infty} p_{C}(r) P( n | \lambda, r) dr,\\
P_{D}(n)&=& \int_{0}^{\infty} p_{D}(r) P( n | \lambda, r) dr,\nonumber\\
P_{T}(n)&=& \int_{0}^{\infty} p_{T}(r) P( n | \lambda, r) dr.\nonumber
\end{eqnarray}
Here $P(n | \lambda, r)$ is the probability distribution for the length $n$ of an intergenic region,  given the interaction range $r$ and the pressure $\lambda$; it depends on the the form of the interaction potential. For instance, if the potential is that of the CF model (equation \ref{U}), then $P(n|\lambda, r)$ is given by  equation \ref{final}. Note that we retrieve the original CF model if we insert $p_{C}(r) = \delta(r-2 \tau)$, $p_{D}(r) = \delta(r-2 \pi )$ and $p_{T}(r) = \delta(r-(\tau+\pi))$ in the above integrals.

\begin{figure*}[htbp]
\includegraphics[width=\textwidth]{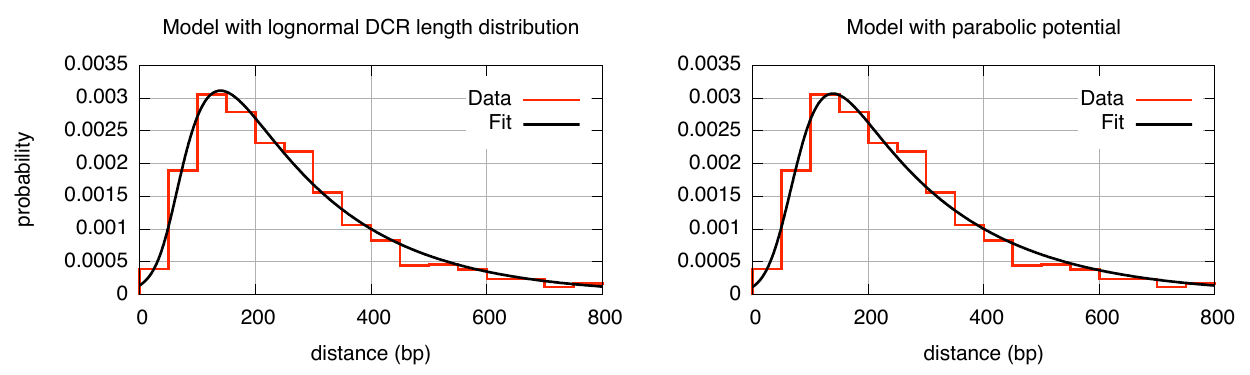}
\caption{\label{figureFits} Fits of two more detailed models to the length distribution of convergent intergenic regions in \emph{S. cerevisiae}. The fact that both models allow for nearly perfect fits shows that one needs additional, independent information to distinguish between the various compatible models. Left: CF model with log-normally distributed DCR lengths. Fit parameters for the log-normal distribution: $\mu=4.61$, $\sigma=0.405$, $\lambda=5.25$, $\epsilon=4.70\times 10^{-2}$. Right: Log-normally distributed DCR lengths and parabolic potential. Fit parameters: $\mu=5.01$, $\sigma=0.314$, $\lambda=4.95\times 10^{-3}$, $U_{0}=4.11$.}
\end{figure*}

In the case of \emph{S. cerevisiae} some studies \cite{David:2006:PNAS, Helden:2000:NA, Graber:2002:NA, Perocchi:2007:NAR} suggest that the distribution of 3'-UTRs is log-normal. We therefore assume that $p_{C}(r)$  is the distribution of the sum of two numbers drawn independently from a log-normal distribution. A sum of log-normally distributed random variables can be approximated reasonably by another log-normal distribution. We therefore assume that $p_{C}(r)$ is log-normal as well (with parameters $\mu$ and $\sigma$). The corresponding fit to the histogram of convergent intergenic distances in \emph{S. cerevisiae} is better than the fit of the CF model and does not show the artifactual sharp peak (see Fig. \ref{figureFits}).
Nevertheless,  the mean of the best-fitting log-normal distribution ($\mu=4.61$, $\sigma=0.405$, $\textrm{mean}=e^{\mu + \sigma^{2}/2}=109$) is rather close to the estimate resulting from the CF model. ($2 \tau=122$). Below we show that a better agreement with experiment can be obtained if we allow for an alternative interaction potential.

\subsection{Alternative potentials}\label{IIB}

In the CF model, we used the simple potential defined in equation \ref{U}. This potential was convenient because of its simplicity (only one parameter) and its straightforward interpretation. It is, however, possible to generalize our approach to alternative potentials. Equation \ref{plxfinal} holds for any finite-ranged potential $U(n)$; this means that equations  \ref{plxfinal} and \ref{integrals} can be used to compute the ORF spacing for arbitrary finite-range potentials.

\subsection{Yeast DCRs}\label{IIC}

In the main text, we mentioned that the CF model predictions for the DCR length are on the low side.  Not much is known about termination sequences in \emph{S. cerevisiae}, but most of the poly-adenylation signals seem to occur within 70 bp from the stop codon \cite{Helden:2000:NA}. Estimates for the median 3' UTR length in \emph{S. cerevisiae} range from 80 to about 100 bp\cite{David:2006:PNAS, Graber:2002:NA}. This shows that, although the CF model does predict the qualitative features of the distributions in Yeast, such as the exponential tail of the distribution, in order to get accurate quantitative agreement with the DCR lengths found in recent experiments, the assumptions of the CF model are too crude. Using the above techniques, we can refine the model and get better agreement.

First, recent studies suggest that the 3'-UTRs in Yeast can be approximated by a log-normal distribution; we therefore now choose $p_{C}(r)$ to be log-normal (with parameters $\mu$ and $\sigma$). Second, recent experiments strongly suggest that many 3'-UTRs are long and that they often overlap considerably \cite{David:2006:PNAS, Perocchi:2007:NAR}; nevertheless, the ORFs hardly ever get closer together than 120 bp. This suggests a model in which the force is not constant; instead, the repulsion seems to be high at short distances, but low at longer distances. One way to model this is to use the following quadratic potential instead of equation \ref{U}:
\begin{equation}\label{U2}
\frac{U(n)}{kT} = \left \{
\begin{array}{ll}
U_{0}(1-\frac{n}{r})^{2} & (n < r)\\
0 & (n \geq r).\\
\end{array}
\right.
\end{equation} 
The fit of this model (with $\mu$, $ \sigma$, $U_{0}$ and $\lambda$ as parameters, but a given mean distance) to the convergent data is excellent (see Fig.  \ref{figureFits}); also, the resulting log-normal distribution for $p_{C}(r)$ has a mean  $e^{\mu + \sigma^{2}/2}=160$, which leads to a mean DCR length of about 80 bps. This is good agreement with the experimental results.

In the study of Van Helden \emph{et al} \cite{Helden:2000:NA}, poly-adenylation signals were found at about 35 bp and 55 bp downstream of the stop codon of ORFs. It is tempting to speculate that these sequences are responsible for the strong repulsion starting at a distance of about 120 bp in convergent intergenic regions.

\subsection{Higher eukaryotes}\label{IID}

In Fig. \ref{fig:CDTvarious} and S\ref{fig:fitall} the intergenic distance distributions for various different organisms are shown. Strikingly, the simple CF model can very well describe the qualitative features of all these model organisms, such as the exponential tail of the distributions and the dependence of the distributions on orientation.

In complex, multicellular eukaryotes, control regions typically are very long and exhibit a high variance( see e.g. \cite{Hajarnavis:2004:NA}). As the lengths and variances increase, the assumptions of the constant force model become less justified. Above we have shown that the CF model can be extended to incorporate alternative potentials and polydisperse interaction ranges. This allows us to produce excellent fits to the data for all organisms. Nevertheless, when it comes to \emph{predicting} the length distributions of UCRs and DCRs for higher organisms, the results depend too sensitively on the choice of the potential to produce meaningful predictions. Therefore we refrain from using the fit parameters for \emph{D. melanogaster}, \emph{ A. thaliana}, \emph{C. elegans} and \emph{P. falciparum} as predictions for the DCR and UCR lengths

\section{Supplementary figures S1 and S2 and fitting procedure}

\begin{figure*}[htbp]

\includegraphics[width=13cm]{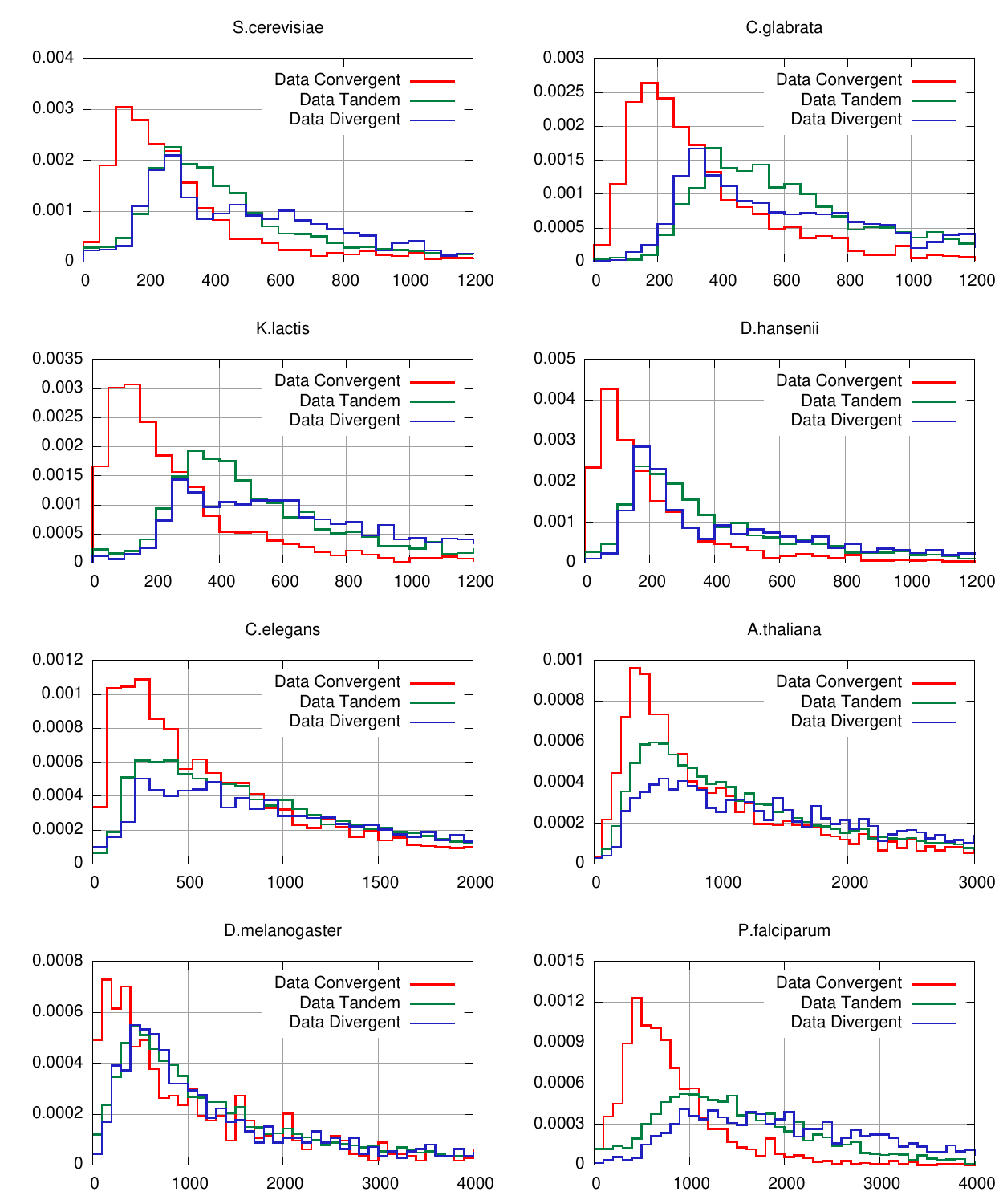}

\caption{\label{fig:CDTvarious} Distributions of intergenic distances, broken down into three different subsets (convergent, tandem, divergent pairs), for four fungi and four additional eukaryotes. Consistently, the convergent gene pairs are, on average, closer together than the tandem ones. The divergent genes are furthest apart. Note also that the divergent distribution is bimodal for all fungi, suggesting the presence of bi-directional promoters in all of them.}
\end{figure*}

\begin{figure*}[htbp]
\includegraphics[width=13cm]{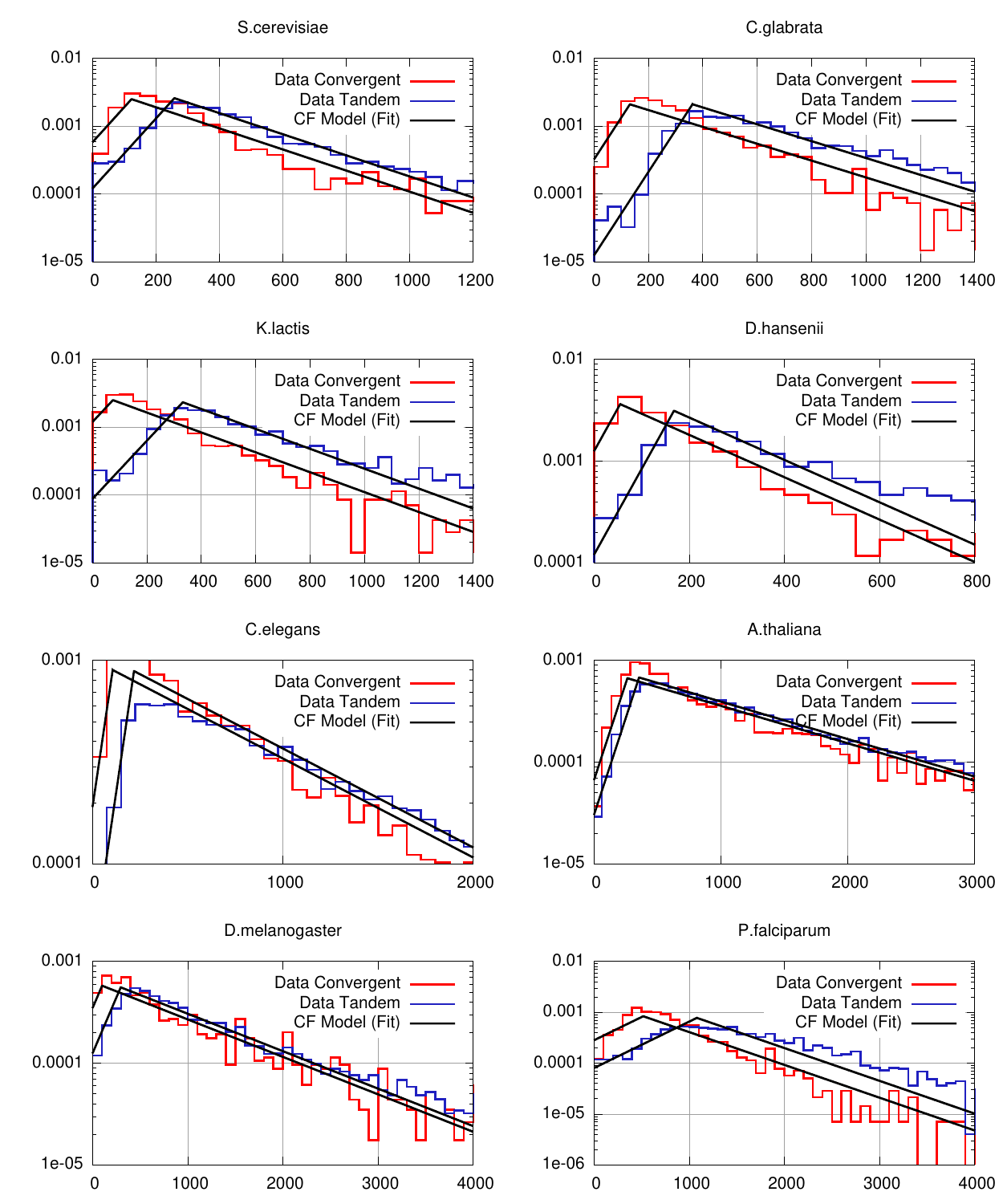}
\caption{\label{fig:fitall} Maximum likelihood fits of the CF model to the distance distributions of four fungi and four additional eukaryotes. Despite the simplicity of the CF model, it does capture the qualitative features of each of the genomes, such as the dependence of the ORF spacing on relative orientation, and the exponential tails. The fits are used to estimate the sizes of UCRs and DCRs for the fungi; see Table 1 in the main text. Such quantitative estimates are probably not reliable for the higher eukaryotes, for which the model assumptions may be too crude.}
\end{figure*}

Fig. \ref{fig:CDTvarious} shows the distributions of intergenic distances for four different fungi and four additional eukaryotes, broken down into three different subsets (convergent, tandem, divergent). The C and T distributions are also displayed in Fig. \ref{fig:fitall} in log-linear scale, combined with fits of the CF model. In case of the fungi, we used these fits to estimate UCR and DCR sizes in these species. The fit parameters are given in Table 1 in the main text.

We used the maximum likelihood method to fit our model to the data and to determine the errors in the fit parameters. For a given set of observed intergenic distances ( $\{n_{C}\}$ and $\{n_{T}\}$ for convergent and tandem pairs, respectively), Bayes' rule states that the likelihood of a set of fit parameters obeys
\begin{eqnarray}
\lefteqn{P( \pi, \tau,  \lambda, \epsilon \left |  \{n_{C}\}, \{n_{T}\}\right) = }\\&& P(\left. \{n_{C}\}, \{n_{T}\}  \right | \pi, \tau, \lambda, \epsilon)
\frac{P(\pi, \tau, \lambda, \epsilon)}{P(\{n_{C}\}, \{n_{T}\})}.\nonumber
\end{eqnarray}
Here  $P(\pi, \tau, \lambda, \epsilon)$ is the prior probability distribution, which we take to be uniform. In that case
\begin{eqnarray}
\lefteqn{P( \pi, \tau, \lambda, \epsilon | \{n_{C}\}, \{n_{T}\})}\\
 & &\propto  P(\left.\{n_{C}\}, \{n_{T}\} \right |  \pi, \tau, \lambda, \epsilon).\nonumber
\end{eqnarray}
The parameter values with maximal likelihood are therefore those that maximize $ P(\left. \{n_{C}\}, \{n_{T}\} \right |  \pi, \tau, \lambda, \epsilon)$. 

In practice, it is more convenient to work with the logarithm of the likelihood, as
\begin{eqnarray}
\lefteqn{\log\left(P(\{n_{C}\}, \{n_{T}\}  |  \pi, \tau, \lambda, \epsilon)\right)} \\
& & = \log \left(\prod_{n\in\{n_{C}\} } P_{C}(n) \prod_{n'\in\{n_{T}\}} P_{T}(n')\right) \nonumber\\
& & = \sum_{n\in\{n_{C}\}} \log\left(P_{C}(n)\right) + \sum_{n' \in\{n_{T}\}} \log \left(P_{T}(n')\right).\nonumber
\end{eqnarray}
If we define 
\begin{eqnarray*}
X_{C}^{\le}&\equiv&\sum_{n\in\{n_{C}\}, n\le2\tau} n,\\
X_{C}^{>}&\equiv&\sum_{n\in\{n_{C}\}, n\ > 2\tau} n,\\
X_{T}^{\le}&\equiv&\sum_{n\in\{n_{T}\}, n \le \tau + \pi} n,\\
X_{T}^{>}&\equiv&\sum_{n\in\{n_{T}\}, n > \tau+\pi} n.
\end{eqnarray*}
and call the total number of convergent and tandem pairs $N_{C}$ and $N_{T}$, this reduces to
\begin{eqnarray}\label{loglike}
\lefteqn{\log(P(\{n_{C}\}, \{n_{T}\}  |  \pi, \tau, \lambda, \epsilon)) } \\
&& = N_{C} \log(c_{1}) + N_{T} \log(c_{2})\nonumber\\
&&\mbox{} - (\lambda-\epsilon)(X_{C}^{\le}+X_{T}^{\le})\nonumber\\
&&\mbox{}- \lambda(X_{C}^{>}+X_{T}^{>})\nonumber\\
 &&\mbox{}+ 2 N_{C} \epsilon \tau  + N_{T} \epsilon (\tau+ \pi),\nonumber
\end{eqnarray}
which can be maximized straightforwardly. To avoid possible influences of rare outliers, we only used values of $n$ that fall in the domain that is plotted Fig. \ref{fig:fitall}. This is correct if we modify $c_{1}$ and $c_{2}$ in equation \ref{loglike} such that, given the domain $D$,
\begin{equation}
\int_{D} P_{C}(n) dn = \int_{D}P_{T}(n) dn =1.
\end{equation} 

If we plot the likelihood as a function of one of the parameters, while keeping the other parameters at their maximum likelihood value, the plot can very well be approximated by a Gaussian. We use the standard deviation of this Gaussian as the error in the maximum likelihood parameter values.

In the main text, we discussed the values of $\tau$ and $\pi$, but not of $q$. The probability that two randomly chosen base pairs are the same and could therefore overlap is $\frac{1}{4}$. The fact that $q$ is much higher than $\frac{1}{4}$ shows that ``overlap'' is much easier than expected based on this argument. This could reflect the density of functional elements, but also the flexibility of functional sequences, and the fact that regulatory regions are not mono-disperse.
 
\section{Statistical tests}

In this section we describe the statistical tests that are mentioned in the main text.

\subsection{Co-expressed divergent pairs in  \emph{S. cerevisiae} are closer together than expected}\label{IVA}

In the main text, we state that co-expressed divergent gene pairs in \emph{S. cerevisiae} have a tendency to have short intergenic regions. We tested this hypothesis as follows.

\begin{figure}[tpb]
\includegraphics[width=8.5cm]{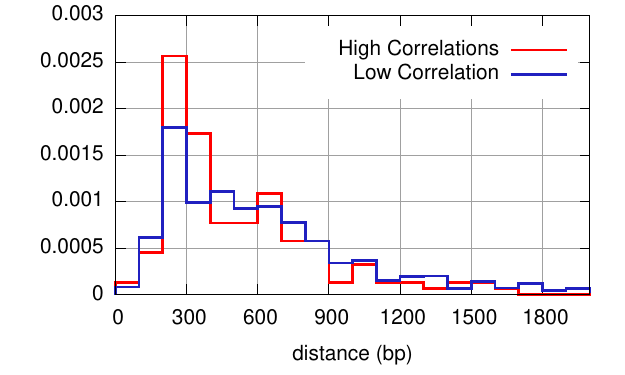}
\caption{\label{fig:highlow}Distance distributions for divergent, neighboring gene pairs with high or low correlation in their expression, in \emph{S. cerevisiae}. The figure shows that pairs with a high correlations coefficient are more likely to be close together than the ones with low correlation coefficients. This fact supports the hypothesis that the pairs in the first peak have a shared, bidirectional \emph{cis}-regulatory region.}
\end{figure}

We used the expression data compiled by Dr. Andre Boorsma and Prof. Harmen J. Bussemaker to compute, for each neighboring pair of genes, the Pearson's correlation coefficient of their expression in about 900 experiments (see references \cite{Boer:2003:BC,Boorsma:2004:Y,Bro:2003:BC,Chu:1998:S,Daran-Lapujade:2004:BC,Devaux:2001:E,Fleming:2002:PNA,Gasch:2000:BC,Gasch:2001:BC,Harris:2001:CB,Hughes:2000:C,Lagorce:2003:BC,McCammon:2003:BC,Mnaimneh:2004:C,Murata:2003:BC,Sahara:2002:BC,Spellman:1998:BC,Tai:2005:BC,Yoshimoto:2002:BC} for the original publications). These coefficients were usually low ($<0.3$). We then split the set of divergent pairs into two subsets: those with a low correlation coefficient ($<0.3$), called set 1, and those with a high one ($\geq 0.3$), called set 2. Next, we used a rank sum test to check whether the intergenic regions in set 2 are indeed shorter than expected at random. 

The rank sum test was performed as follows. We first ranked the intergenic regions according to their length. Then, we computed the sum of the ranks of the intergenic regions in set 2; we call it $R$. Next, we randomized the ranks in the data set $10^{7}$ times, each time computing the rank sum of set 2 in the randomized data. Finally, we counted the number of times $R$ was smaller or equal to the rank sums obtained from the randomized data. The results show that the ORF pairs with a high correlation coefficient are significantly closer together than the ones with a low one ($p<0.002$). 

We checked that the observed signal is not due to paralogous gene pairs by excluding them from the set and repeating the test; this did not change the result. As a control experiment, we  tested whether the same signal is also present in the set of \emph{tandem} neighbors. This is not the case (rank sum test: $p=0.56$). We note, however, that a similar signal \emph{was} present in the set of convergent pairs ($p<2\times 10^{-4}$); we have no satisfactory explanation for this fact.

The mean intergenic distance in set 1 is 721 bp; in set 2, it is 558 bp. The difference in the distance distributions of both sets is visually apparent (see Fig. \ref{fig:highlow}). 

\subsection{Divergent genes in \emph{S. cerevisiae} are more likely to be associated with the same process or component if they are close together }\label{secGO}\label{IVB}

We studied whether there is an association between intergenic distance and functional similarity in divergent gene pairs in \emph{S. cerevisiae}. In order to do this, we need to be able to quantify the functional similarity between two given genes. For this purpose we used the GO annotations of the GO Consortium (version 5.463 of Aug. 22 2007, see reference \cite{Ashburner:2000:G}) and the information-theoretic measure for semantic similarity proposed by Resnik  \cite{Resnik:1998:JAIR}.

The Gene Ontology is a hierarchical vocabulary of terms that can be assigned to genes. It falls apart into three independent taxonomies, each defined to describe one aspect of genes: the {\em biological process} they are involved in, the {\em molecular function} they perform and the {\em cellular component} they are active in.

The semantic similarity measure of Resnik provides, for each pair of GO terms $a$ and $b$, a similarity score $s(a,b)$. 
If, for a given aspect $\alpha$ (biological process, molecular function or cellular component),  gene $A$ has been assigned GO terms $a_{1} \ldots a_{n}$, and gene $B$ has been assigned term $b_{1}\ldots b_{m}$, then we define the similarity between genes $A$ and $B$ on this aspect as:
\begin{equation}
S(A, B, \alpha) = \max_{i,j} s(a_{i}, b_{j}).
\end{equation}
Thus we can compute similarity values for each gene pair and for each aspect of the GO ontology.

If a certain gene did not have any assignment for some aspect, then the similarity score with any other gene was considered undefined on this aspect, and the gene was excluded from the analysis corresponding to this aspect.

We performed the following statistical tests. First, we divided the data set in two subsets, set 1 and set 2. The first set consisted of all divergent pairs that were in the ``first'' peak and set 2 contained all divergent neighbors that were further apart. As the bordering value we chose  $d=2 \pi=375$, since intergenic regions that are longer than that value can easily accommodate two independent promoters. We applied a Wilcoxon-Mann-Whitney rank-sum test to challenge the null hypothesis that the similarity scores of the pairs in set 1 and set 2 are drawn from the same distribution. We repeated this test for each aspect of the GO. The test results are $p=8.8\times 10^{-4}$, $p=0.06$ and $p=4.8\times 10^{-5}$ for biological process, molecular function and cellular component respectively. We also ran a Spearman rank correlation test on the data, which resulted in the values $p=2.0\times 10^{-5}$, $p=0.24$ and $p=1.3\times 10^{-5}$. 

We conclude that the similarity scores belonging to the aspects biological process and cellular component are associated with intergenic distance. We do not, however, find a significant association between molecular function and intergenic distance. This is not very surprising, as proteins with a similar molecular function (e.g. ``DNA binding'' proteins), can act in very different processes and cellular components, so that there is no clear a priori reason to co-regulate them using a shared UCR.

We repeated this analysis for the convergent and tandem gene pairs. For the convergent pairs, none of the statistics were significant. However, the tandem pairs showed a similar pattern as the divergent ones; the Spearman rank correlation test resulted in $p=4.6\times10^{-5}$, $p=0.41$ and $p=1.8\times 10^{-5}$ for biological process, molecular function and cellular component respectively.

At this point, it is illustrative to point at one interesting example in Yeast: the BIO3, BIO4 and BIO5 cluster \cite{Phalip:1999:G}. All genes in this cluster are involved in the biotin biosynthesis pathway. BIO3 and BIO4 are transcribed in a divergent orientation from a short intergenic region of length 222 bp (which falls into the first peak in the length distribution of divergent intergenic regions) and are tightly co-expressed. The orthologs of BIO3 and BIO4 in \emph{E. coli} are BioA and BioB; these genes are closely spaced divergent neighbors as well (87 bp), and are simultaneously repressed by BirA binding to their shared UCR.  Phalip \emph{et al.} already speculate that a similar mechanism is at work in \emph{S. cerevisiae} \cite{Phalip:1999:G}, but the mechanism of co-expression of these genes has not been studied in detail. BIO4 and BIO5 are tandem neighbors, and are only 55 bp apart, Clearly, this cluster has many of the features that we see in our statistical analysis. We therefore suggest that detailed experimental work on the regulation of the BIO cluster might illuminate some important mechanisms that shape the distribution of genes over the \emph{S. cerevisiae} chromosome. 

\subsection{Greater than expected number of convergent intergenic regions with length 20-60 bp in  \emph{E. coli}}\label{IVC}

Here we show that the number of convergent intergenic regions with a length in the range 20-60 bp, is significantly larger than expected in \emph{E. coli}.

In the calculation below, we estimate the significance of the peak in a very conservative way. We take the best-fitting exponential probability distribution as our null distribution (the fit is shown in Fig. 1b in the main text; its scale factor equals 145 bp). This way, we \emph{underestimate} the statistical significance of the peak as we ignore the fact that the CF model actually predicts a \emph{dip} in the distribution at the place of the peak.

Given the exponential null distribution, the fraction of the sample that is expected in the domain 20-60 bp is 0.21. Since the total number of convergent pairs is 543, the number of pairs in this domain is a random variable $X$ that is distributed binomially with $p=0.21$ and $N=543$. The observed number of pairs in this domain in \emph{E. coli} is 198; the probability for this to happen given the null distribution is $P(x\geq198; p=0.21, n=543)<10^{-13}$.

Based on the numbers above we should have expected $0.21\times 543=114$ pairs in the domain 20-60 bp. The actual observed number is 198; this means that we need about 86 bi-directional terminators to explain the data. If we assume that \emph{E. coli} has 750 operons, we estimate that at least $\frac{2\times 86}{750}\times 100\%=23\%$ of the operons is terminated by a bi-directional terminator. 

\subsection{In \emph{E. coli}, putative terminators in C regions are more often bi-directional than those in T regions}\label{IVD}

Here we show that the fraction of putative terminators that is classified as bi-directional by RNAMotif software \cite{Lesnik:2001:NAR}, is larger in C regions than in T regions.

The statistical test was performed as follows. Our null hypothesis is that the terminators in the C region are a random sample from the total set of terminators in C or T regions. In total, the C and T regions together contain 1198 putative terminators, of which 222 are classified as bi-directional by RNAMotif. The C regions contain 378 putative terminators, of which 104 are bi-directional according to RNAMotif. If the 378 are chosen at random from the total set of 1198 terminators, then the number of terminators in the sample that are classified as bi-directional is a hypergeometric random variable. The probability to observe at least 104 bi-directional terminators in a random sample of 378 terminators, taken from a set of 1198 terminators containing 222 bi-directional ones, equals $1\times 10^{-7}$.  

\subsection{Putative bi-directional terminators in C regions tend to occur in short regions}\label{IVE}

The fraction of the putative terminators in convergent intergenic regions that could be bi-directional (according to the RNAMotif algorithm) is significantly larger in short intergenic regions ($<100$bp) than in long ones. We used the same statistical test as in the previous subsection. In total, the C regions contain 378 putative terminators; of these, 104 are classified as bi-directional. The short convergent intergenic regions contain 158 putative terminators, of which 62 are bi-directional according to RNAMotif. The probability to observe at least 62 bi-directional terminators in a random sample of 158 terminators, taken from a set of 378 terminators containing 104 bi-directional ones, is $2\times 10^{-5}$.

\subsection{Convergent operons that are close together are not more often functionally related}\label{IVF}

We tested whether the convergent operons that are close together ($<70$bp) are more likely to be active in the same biological process or cellular component than ones that are further apart. For this we used the GO annotations from the GOA Database \cite{Camon:2004:NA}(version date: September 9. 2007). The same method was used as in section \ref{secGO}, except that we now had to perform the analysis on the level of operons rather than genes. In order to compute the similarity between two operons, we compared the GO assignments for each gene in the first operon with each gene in second; the maximum of these scores was used as a similarity measure for the operons.
We did not find a significant signal for any of the aspects of the Gene Ontology ( molecular function: $p=0.20$; biological process: $p=0.25$; cellular component: $p=0.27$).

\bibliographystyle{ieeetr}

\end{document}